\documentclass[preprint,showpacs,preprintnumbers,amsmath,amssymb,showkeys]{revtex4-1}
\usepackage{graphicx}
\usepackage{float}
\usepackage{color}
\usepackage{ulem}

\def\per{$\rm [\bar{1} \bar{1} 2]~$}
\def\para{$\rm [1 \bar{1} 0]~$}

\begin{document}

\title{Observation of correlated spin-orbit order in a strongly anisotropic quantum wire system}

\author{C. Brand$^1$} \author{H. Pfn\"ur$^1$}  \author{G. Landolt$^{2,3}$} \author{S. Muff$^{3,4}$}   \author{J. H. Dil$^{3,4}$} \author{Tanmoy Das$^{5}$} \email{Corresponding author: tnmydas@gmail.com} \author{C. Tegenkamp$^{1}$} \email{Corresponding author: tegenkamp@fkp.uni-hannover.de}
\affiliation{$^1$ Institut f\"ur Festk\"orperphysik, Leibniz Universit\"at
Hannover, Appelstra\ss e 2, 30167 Hannover, Germany}
\affiliation{$^2$ Physik-Institut, Universit\"at Z\"urich, Winterthurerstrasse 190, 8057 Z\"urich,
Switzerland}
\affiliation{$^3$ Swiss Light Source, Paul Scherrer Institute, 5232 Villigen PSI, Switzerland}
\affiliation{$^4$ Institute of Condensed Matter Physics, $\acute{E}$cole Polytechnique $F\acute{e}d\acute{e}$rale de Lausanne, Switzerland}
\affiliation{$^5$ Department of Physics, Indian Institute of Science, Bangalore 560012, India}
\date{\today}
\begin{abstract}
Quantum wires with spin-orbit coupling  provide a unique opportunity to simultaneously control the coupling strength and the screened Coulomb interactions where new exotic phases of matter can be explored. Here we report on the observation of an exotic spin-orbit density wave  in Pb-atomic wires on Si(557) surfaces  by mapping out the evolution of the modulated spin-texture  at various conditions with spin- and angle-resolved photoelectron spectroscopy. The results are independently quantified by surface transport measurements. The spin polarization, coherence length, spin dephasing rate, and the associated quasiparticle gap decrease simultaneously as the screened Coulomb interaction decreases with increasing excess coverage, providing a new mechanism for generating and manipulating a spin-orbit entanglement effect via electronic interaction. Despite clear evidence of spontaneous spin-rotation symmetry breaking and modulation of spin-momentum structure as a function of excess coverage, the average spin-polarization over the Brillouin zone vanishes, indicating that time-reversal symmetry is intact as theoretically predicted.
\end{abstract}
\maketitle


The spin degree of freedom of electrons has attracted much attention over the last several years and plays a central role in spintronic device structures \cite{Sinova12}, the spin Hall effect \cite{Kato04,Koenig07}, topological insulators \cite{Hsieh08}, Rashba-type surface or interface states \cite{Rojas13,Santander14} and others. Typically the propagation of electrons is treated as a single particle problem. This simple concept fails, however, when electronic correlation effects become important and spin- and orbit-effects are entangled like in hard-magnets \cite{Larson03},
$ j$=1/2 Mott-insulators \cite{Kim08} or in new emergent topological superconductors  \cite{Hor10,Yang12,Das12b}.
Among others, the spin-ordered Au-chain ensemble grown on Si(553) \cite{Barke06,Erwin10,Aulbach13} and also magnetic chain structures on insulating or superconducting surfaces \cite{Loth12,Gauyacq13,Kim14} are recent illustrations of this entanglement and have attracted a lot of research activities due to their full flexibility for manipulation with atomic precision.

Reminiscent of antiferromagnetism in Cr \cite{Overhauser62}, the problem of spin-orbit coupling (SOC) under consideration of electronic interaction has been revisited recently \cite{Das12,Das12b}. It was shown that inherent spin splitting such as Rashba-type SOC yields Fermi surfaces with nesting between opposite helical states causing a so-called spin-orbit density wave (SODW) state which cannot be characterized by independent local order parameters. This new emergent
phase of matter arises without breaking time-reversal symmetry above a critical value for the Coulomb potential $U$ and its order parameter, i.e. the energy gap $\Delta $, is determined by the energy scales of both the SOC strength $\lambda$ and the interaction $U$ \cite{Das12}. Furthermore, a finite gap $\Delta$ protects the SODW from spin dephasing against  external perturbations like magnetic fields, thermal excitation and doping by excess coverage. In case that the interaction energy
is large compared to the SOC, however, the SODW can be  overturned at the expense of a spin density wave (SDW).

Experimentally a SODW can be identified by the difference between charge and spin order, the nesting of spin polarized states and the concomitant opening of a gap, and the depolarization of the spin order as a function of Coulomb screening due to the dephasing of spin states. The observation of all these signatures in our experimental data will be addressed subsequently after a brief introduction into our quantum wire system of choice, i.e. epitaxially grown Pb-wires on Si(557) \cite{Czubanowski07,Tegenkamp07,Kim07}.

\begin{figure}[tb]
\begin{center}
\includegraphics[width=0.7\columnwidth]{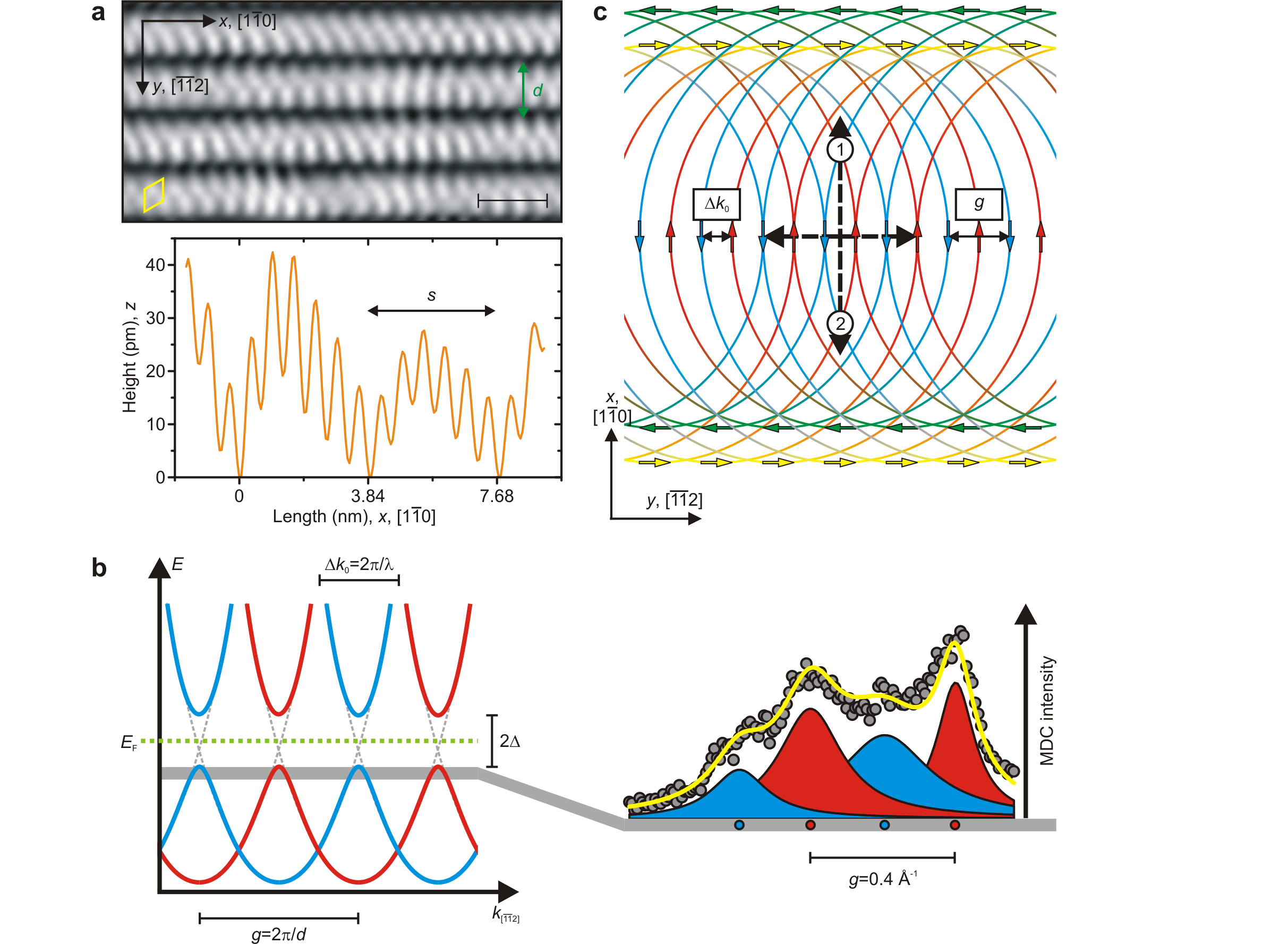}
\caption{\label{FIG1} {}}
\end{center} \end{figure}

Most remarkably, at the critical concentration of 1.31 monolayers (ML), a one-dimensional gap $\Delta$=20~meV  is opened below 78~K so that the Pb layer becomes insulating in the direction perpendicular to the wires while it remains conducting along the chain direction \cite{Tegenkamp05,Tegenkamp08}.
As sketched in Fig.~\ref{FIG1}b, this band gap opening is caused by Pb-induced
refacetting of the (557) orientation to (223) oriented facets, thus introducing a
new reciprocal lattice vector, $\rm {\bf g}$= $2\pi d^{-1}$ $\rm {\bf e}$, which fulfils the (spin-polarized) nesting condition $\rm 2 {\bf k_F = g}$ along the \per direction, which is one of the signatures for the formation of a SODW as discussed above ($d$ is the interwire distance, $\rm \bf k_F$ the Fermi wave vector, $\rm \bf e$ is the unit vector).
This condition is superimposed on the local 2D character of the densely packed
 $\alpha-\rm \sqrt 3 \times \sqrt3$ phase of Pb on the mini-(111) terraces,
which is reflected by closed Fermi circles rather than by open Fermi lines (Fig.~\ref{FIG1}c).
The Fermi nesting in this system leads to the expected renormalization of the Fermi surface \cite{Tegenkamp08}. For the sake of simplicity and the fact that the gap is comparably small we show here the quasi-Fermi surfaces, i.e. the circularly shaped spectroscopic features seen at $E_\text{F}$ - 50 meV. This representation takes also the replica bands into account which are seen by the MDCs
taken necessarily also slightly below $E_\text{F}$.
A large degree of geometrical order along the wires, coupled with high 1D conductance, is evident by the 10-fold superperiodicity $(s=$10$\times a_\text{Si})$ at the critical coverage of 1.31~ML, shown by high resolution scanning tunneling microscopy (STM) in Fig.~\ref{FIG1}a and also seen by spot splitting in low energy electron diffraction (LEED, see below).
Strikingly, the system reveals giant Rashba-splitting of $\Delta k_0$=0.2~\AA$^{-1}$ (Rashba parameter $\alpha_\text{R}$=1.9~eV \AA) \cite{Tegenkamp12} which is by far larger than the expected value for chemisorbed Pb monolayers on isotropic semiconducting surfaces, e.g. Pb/Ge(111) \cite{Yaji10}. $\Delta k_0$
corresponds to exactly $\rm {\bf g}/2$, i.e. the periodicity of spin ordering is twice than that of charge ordering
at this critical Pb concentration, so that spins on adjacent terraces are coupled in an anti-parallel manner.
This spin dependent splitting is marked in Figs.~\ref{FIG1}b,c by different colors. The different periodicities of spin and charge are a necessary prerequisite
for non-trivial coupling between them.

Here we show that in this array of atomic wires bound to the (223) oriented  facets on Si(557) we have the unique possibility to tailor the interwire coupling by adsorption of minute amounts of excess coverage $\delta \Theta$.
The screened Coulomb potential scales with the size of the excess coverage as
$ U(\delta \Theta) = U(0)\cdot exp (-q \delta \Theta)$, where $q$  is the screening parameter \cite{Tegenkamp10}.
As we will show, while the spin-orbit order is preserved (at least) up to a critical additional coverage of 0.1~ML, steps are decorated by this excess coverage not randomly, but by formation of superstructures whose periodicity depends on excess concentration of Pb \cite{Czubanowski07,Czubanowski09}. This highlights that the mechanisms of spin and charge order are not the same. The coupling between charge and spin order  is mediated by 1D in-gap states and is reduced
with increasing excess coverage. This leads to
spin depolarization as a function of excess coverage. Therefore this mechanism has much more fundamental consequences than the band shift associated with standard doping, an observation that has been made also in other metallic chain
systems on surfaces \cite{Kirakosian03,Krieg14}, however with somewhat different physical
scenarios. A collapse of the SODW phase is associated by an almost vanishing spin polarization close to  $\delta \Theta \approx$0.2~ML, where the Pb wires turn into an anisotropic 2D metal. Quite naturally, SODW formation and the reduced spin-charge coupling as a function of excess Pb concentration also influence other physical properties that are sensitive to both charge and spin, like
magnetotransport. As expected, the gradual spin depolarization seen in photoemission is closely  related to the spin dephasing probed by magnetotransport measurements \cite{Luekermann10}. This relationship can be quantitatively described by SODW theory, as we will demonstrate below.

\begin{figure}[tb]
\begin{center}
\includegraphics[width=\columnwidth]{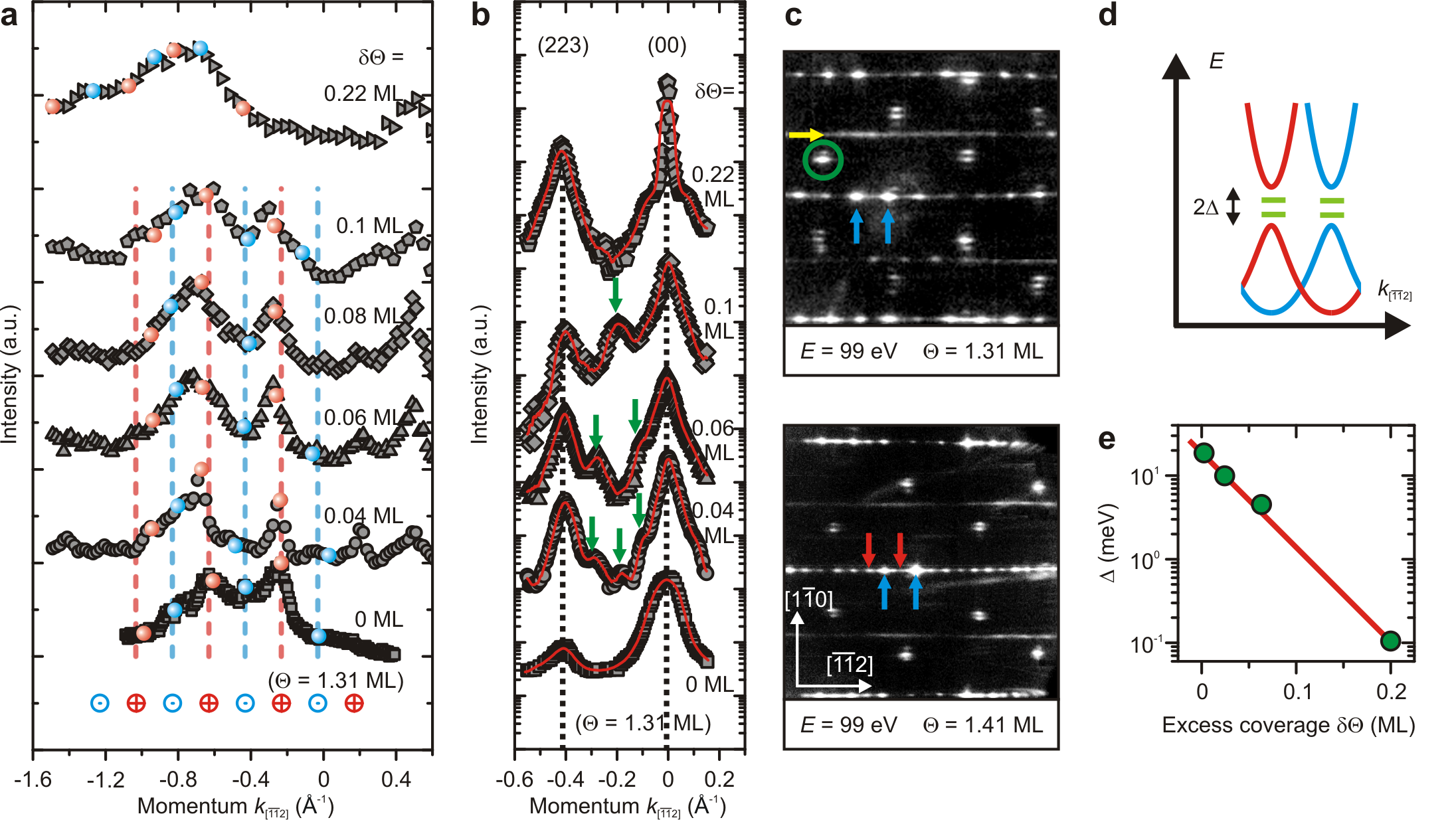}
\caption{\label{FIG2} {}}
\end{center} \end{figure}

\textbf{Results}\\
\textbf{The effect of excess coverage.} Adsorption of excess coverage on the perfect Pb-chain system, as shown in Fig.~\ref{FIG1}a, changes primarily the  periodicities of the Si(557) surface perpendicular to the wires. We demonstrate by the results collected in Fig.~\ref{FIG2} that spin and geometric order indeed do not have the same periodicities, although they are connected and commensurate to each other by the underlying terrace structure. Moreover, while the charge order follows the geometry, the spin order turns out to be robust and unchanged by excess coverage.

The respective charge and spin order can be seen in Fig.~\ref{FIG2}a which shows
the electronic structure perpendicular to the wires. Here the periodicity of the facet structure is  seen in spectroscopy by Umklapp scattering.
From the intense momentum distribution curve (MDC) peak structure of the perfect wire ensemble (Fig.~\ref{FIG1}b) a modulation vector of $\rm 0.4~\AA^{-1}$ can be deduced in agreement with the spot splitting along the \per
direction visible in LEED.  However, as mentioned before, the MDC signal reveals an additional substructure,
which is shifted by  $\rm 0.2~\AA^{-1}$ and has the opposite
spin helicity \cite{Tegenkamp12}. In Fig.~\ref{FIG2}a we marked the extrema of
the $x$-component of the spin polarization by red and blue colored circles. This equidistant splitting is a spectroscopic hallmark for a SODW in its ground state \cite{Das12}.
Note that  we present throughout the paper spin-integrated MDCs measured with high resolution.
The color-coded spin components shown are deduced from additional MDCs recorded with the
Mott detector (see Supplementary Note 1 and Supplementary Figure 1).  In order to reveal good signal/noise ratios in reasonable time the spin distribution has been measured 50~meV below the Fermi level. However, the excellent agreement with results from DC- and magnetotransport demonstrate that this limitation does not affect our conclusions at all.

With increasing $\delta\Theta$, we see that the intensities in the MDCs are slightly altered
due to changing photoelectron scattering conditions related to the geometrical superperiodicities
(see below).
However, and more importantly, the positions of the maxima of spin polarization remain
unchanged in k-space. In contrast to the geometric order, the original spin periodicity
found at $\delta\Theta$=0~ML of 0.2 and 0.4 \AA$^{-1}$, respectively, prevails until we reach
the limit of second Pb layer growth at an excess coverage of 0.2~ML.
At this excess coverage also the original SODW phase collapses, as seen by negligible spin
polarization of all three components.
Since the enhanced spin-texture along the Fermi surface (FS) nesting direction only appears below 0.2 ML excess coverage and below 78~K, it strongly indicates to be an interaction effect, rather than a single particle SOC effect. We therefore conclude that the periodicity of spin order up to $\delta\Theta$=0.2~ML is exclusively determined by the periodicity of the terrace structure of (223) facets.
This terrace induced spin order is reached with maximum spin polarization at the critical
concentration of $\Theta$=1.31~ML, and is interpreted to represent the ground state of the SODW.

Looking at the geometric order, the diffraction pattern of the perfect phase ($ \delta \Theta$=0~ML, Fig.~\ref{FIG2}c) reveals several characteristic features, e.g. pronounced facet spots along \per (blue arrows), $\rm \sqrt3 \times \sqrt 3$ spots (green circle) and intensity streaks at $\rm \times 2$-positions (yellow arrows) originating from the long range ordered (223) facet structure (step density 21~\%), densely packed Pb-reconstruction on the (111)-units, and Si-dimerization taking place at step sites, respectively. In
high resolution LEED studies the $\rm \sqrt3 \times \sqrt 3$ spots  show  splitting of $2\pi/s$=0.16~\AA$^{-1}$ due to formation of domain walls giving rise to 10-fold periodicity
($s$=38.4~\AA) along the wires \cite{Czubanowski07,Czubanowski09}, as confirmed
by our STM study (Fig.~\ref{FIG1}a, Supplementary Note 2, Supplementary Figures 4,5). Similar observations were made for
Pb/Si(553) \cite{Kopciuszynski13}. Upon deposition of excess coverage, superstructure spots appear
within the initial diffraction pattern of the 1.31~ML phase whose periodicity is reduced
by integer multiples of the terrace lattice constant as a function of $\delta\Theta$ (see diffraction line profiles perpendicular to the steps plotted in Fig.~\ref{FIG2}b).
As an example, for  $\delta \Theta$=0.1~ML the periodicity along the \per direction
is doubled (see red arrows) by decoration of every second terrace \cite{Czubanowski09}.

This variation of geometric order induced by excess coverage is coupled
with the reduction of the  band gap, as measured previously  \cite{Tegenkamp10}.
The results, reproduced in Fig.~\ref{FIG2}e show that subbands are formed within the band gap (see schematic of Fig.~\ref{FIG2}d), which gradually fill the gap as a function of $\delta\Theta$. The gradual decrease of the band gap follows $\Delta=20~\text{meV}\cdot exp(-\delta \Theta /0.038~\text{ML})$, which corresponds to $q$=26.3 ML$^{-1}$.
Regarding the influence of geometric periodicities on the spin order observed,
there is obviously no direct impact of the modified chain structures, but the (223) structure
is still observed in LEED at all excess coverages and at low temperature (cf. Fig.~\ref{FIG2}b).
This is the only geometric feature remaining constant upon adsorption of excess coverage.
Therefore coupling between spins on adjacent terraces can only be mediated via the  band gap.
Consequently, the reduction of the band gap also reduces the effective coupling
between spin order on different terraces. In other words, decoration of the step edges with various
periodicities and excess coverages increases  the screened Coulomb interaction,
which couples to spin order via spin-orbit interaction. This is exactly the SODW scenario.

\begin{figure}[tb]
\begin{center}
\includegraphics[width=\columnwidth]{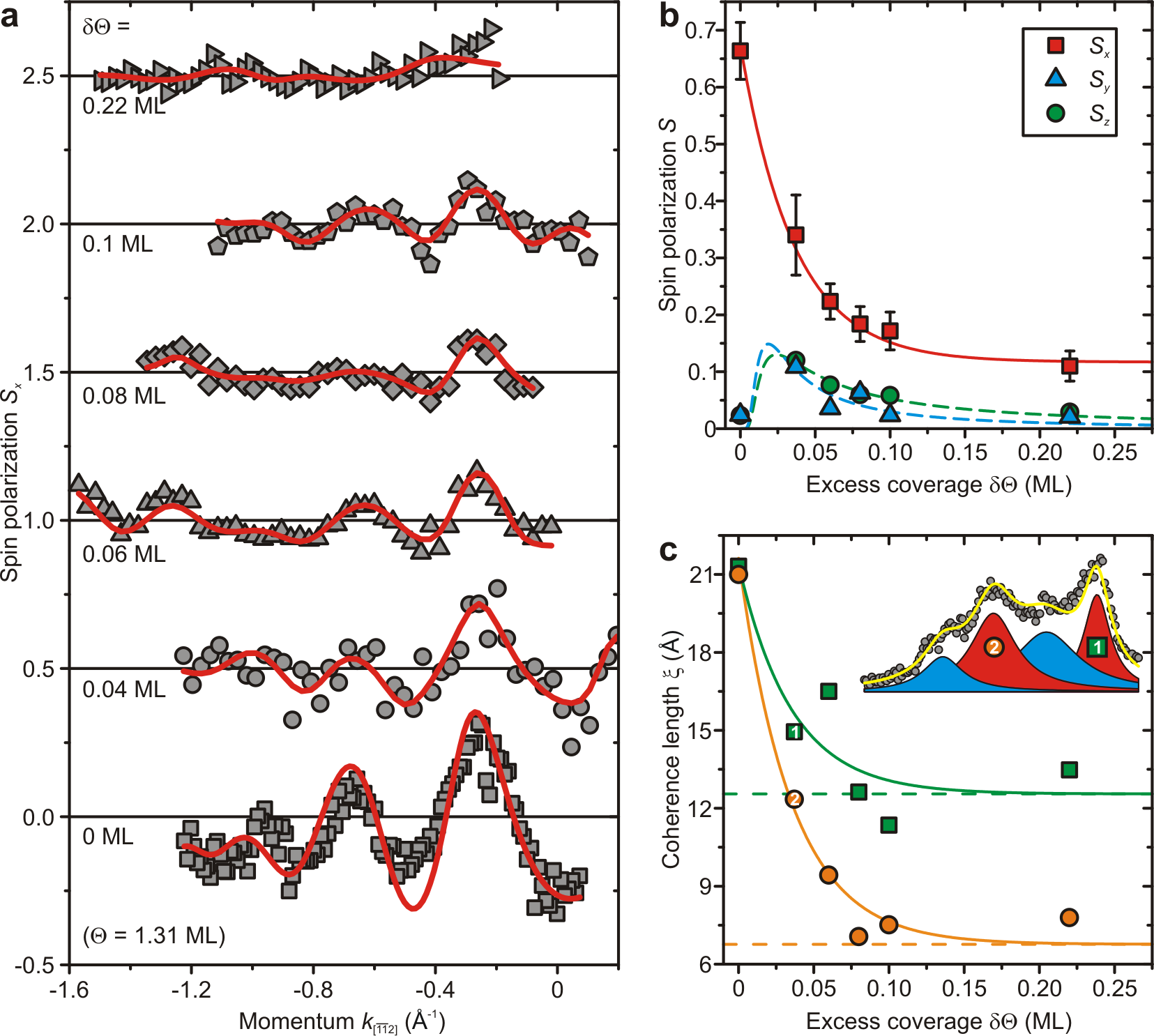}
\caption{\label{FIG3} {}}
\end{center} \end{figure}

\textbf{Spin dephasing in ARPES.} Whether this concept holds can be tested by looking at excitations of the ground state represented by the spin order at the critical Pb concentration of 1.31 ML. Regardless of the protection of the SODW phase the spin polarization of the surface bands should be affected by the excess coverage since the spin dephasing time is proportional to the size of the electronic gap \cite{Das12}.
Fig.~\ref{FIG3}a shows $S_x$-polarization curves obtained along the \per direction, revealing a pronounced spin polarization along the wires (\para direction). Up to $\delta \Theta$=0.1~ML the polarization curves reveal an harmonic but damped oscillatory behavior. The period of $\rm 0.4~\AA^{-1}$ as well as the almost symmetric amplitudes, is giving rise to a vanishing net-polarization within the Brillouin zone ($\rm {\bf P}$=0), indicating the presence of time-reversal symmetry. This observation excludes the possibility of a SDW and further favors the SODW scenario \cite{Das12}.

In order to quantify the spin depolarization we have plotted the peak-to-peak values of the most intense oscillations for all three spin components (Fig.~\ref{FIG3}b). The $S_x$-component decreases exponentially with increasing excess coverage and can be well described by $S_x(\delta \Theta)=0.56 \cdot exp (-\delta \Theta / 0.036~\text{ML})$ + 0.12, i.e. $q$=27.8~ ML$^{-1}$. The (solid) line is a fit according to the SODW theory (see below).
We want to emphasize that the gradual decrease of the band gap $\Delta$ deduced independently from DC-transport measurements reveals an almost identical $q$-value.
As shown in Fig.~\ref{FIG3}b the $S_y$- and $S_z$- components are considerably weaker (see also Supplementary Figure 2). For an ideal Rashba system these components should vanish, while in presence of excess coverage this rule is lifted and might be indicative of spin spirals \cite{Das12b}.
After the band gap has vanished a fast dephasing of the spin-polarized photoholes within the SODW state  mimics almost unpolarized surface bands. As expected, the effect of depolarization comes along with a reduction of the coherence length $\xi$. Fig.~\ref{FIG3}c shows $\xi$ of the photoelectrons deduced from the full width of half maximum  FWHM=2$\xi^{-1}$ for two MDC peaks as a function of excess coverage which follow a similar trend as the spin polarization.

\begin{figure}[tb]
\begin{center}
\includegraphics[width=\columnwidth]{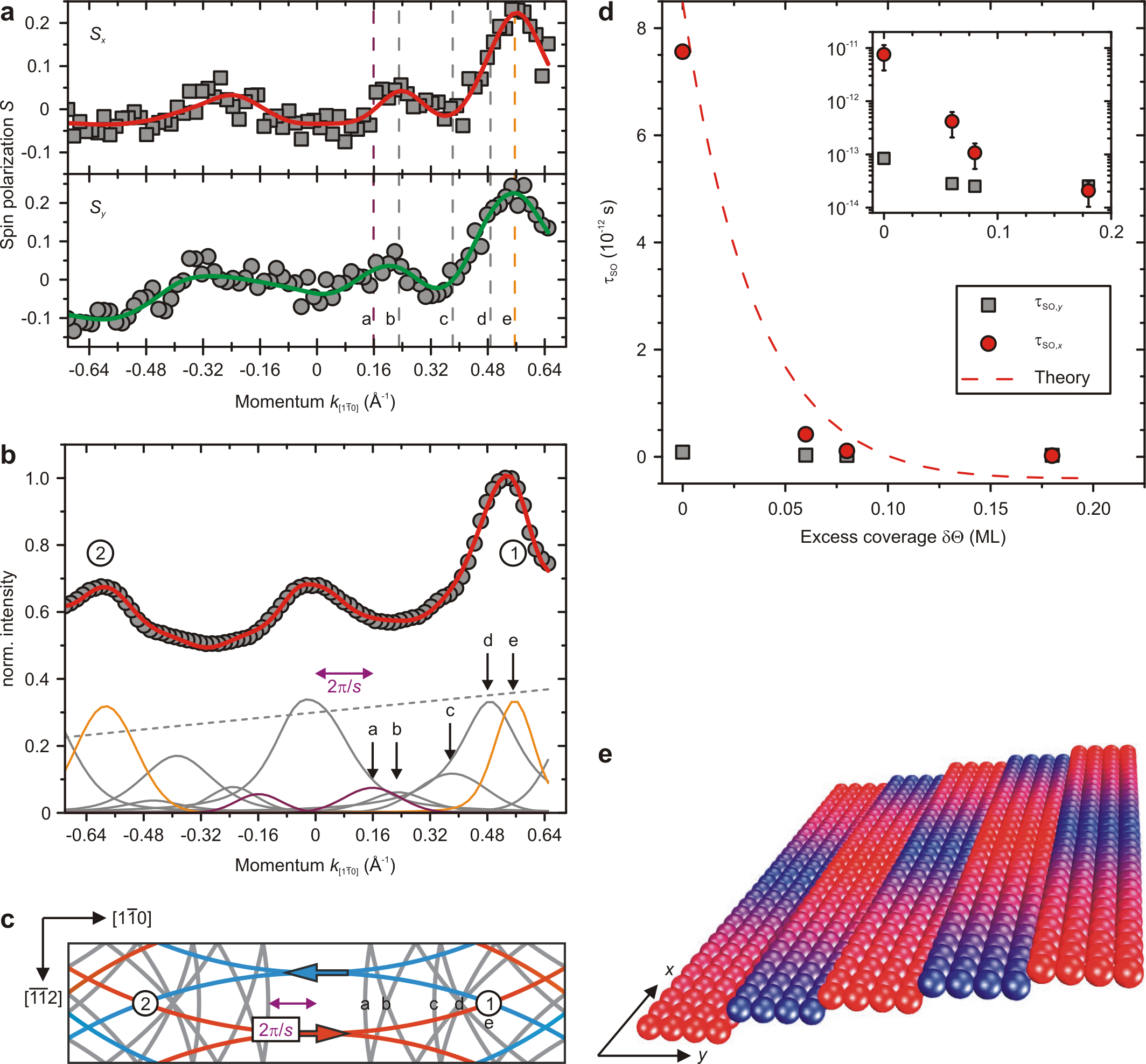}
\caption{\label{FIG4} {}}
\end{center} \end{figure}

Moreover, the spin texture along the wires for the quasi-perfect structure has also been measured. As obvious from the MDC shown in Fig.~\ref{FIG4}b, the most intense features stem from Fermi circles around  $ \overline{\Gamma}$--points of adjacent Brillouin zones and their replicas as depicted by the corresponding labels in Fig.~\ref{FIG4}c. In agreement with the analysis of transport presented below only the central part of the momentum space is relevant. The $2\pi / s$-modulation reflecting the domain wall configuration along the wires (cf. Ref. \cite{Tegenkamp08}) is seen mainly by further satellite peaks shown in Fig.~\ref{FIG4}b. Only their consideration allows a comprehensive description of the spin polarizations and corresponding Mott-MDCs (Fig.~\ref{FIG4}a). At least for the $S_x$- and $S_y$-components significant spin signals are found. Most importantly, the spin signal around $k \approx 0~\AA^{-1}$ reveals only a single spin component giving rise to a spin polarized transport channel as we will show below ($S_z$-component shown in Supplementary Figure 3).

\textbf{Surface magnetotransport.} The dephasing behavior can also be derived from magneto conductance, mainly stemming from the electronic states at $E_\text{F}$ \cite{Luekermann10}. The scattering times for various amounts of excess coverage have been determined from a Hikami analysis \cite{Hikami80} of the magnetotransport data and are plotted in Fig.~\ref{FIG4}d (for further details see Supplementary Note 3 and Supplementary Figure 6). In context of the depolarization presented above, the spin-orbit scattering time $\rm \tau_{so}$ is of particular interest.
As obvious from Fig.~\ref{FIG4}d, the spin-orbit scattering rate deduced for the \para direction ($x$, along the wires) depends crucially on the amount of excess coverage. The strong suppression of spin-orbit scattering in absence of any excess coverage is reminiscent of spin-polarized transport along the wires. Furthermore, the spin-orbit scattering rate is reduced by at least 2 orders of magnitude upon deposition of excess coverage and reveals a close correlation with the depolarization and decoherence behavior seen in angle resolved photoemission spectroscopy (ARPES). We want to point out that this is \textit{per se} not self-evident because ARPES and transport probe electrons at different energies. Regardless, it is remarkable that the spin polarization follows the same trend which in turn demonstrates the close entanglement of both quantities and is a hallmark of strong electronic correlation. As we will show below, both can be consistently explained in terms of the SODW formalism.

\textbf{Discussion}\\
Based on our findings and the fact that within first order the spin lifetime $\rm \tau_s$ is given by the spin-orbit scattering time $\rm \tau_{so}$ in a strongly spin-orbit coupled system, the spin dynamics can be described by a kinetic equation \cite{Tarasenko06}:
\begin{eqnarray}
\frac{\delta {\bf S_k}}{\delta t}+ {\bf S_k} \times {\bf {\Omega}_k}=\frac{<{\bf S_k}>- {\bf S_k}}{\eta}+{\bf P}
\end{eqnarray}
where $\rm {\bf S_k}$ is the spin polarization vector, $\rm {\bf P}$ is any external spin source, and $\eta$ is the scattering time. $\rm {\bf \Omega_k}$ is the effective Larmor
frequency defined in our case by $\rm {\bf {\Omega}_k}=( \Omega_R {\it k_x /k_\text{F}}, 0, \Omega_{SODW})$, where  $\Omega_\text{R}=2 \alpha_\text{R} k_\text{F}/ \hbar$, and
$\Omega_\text{SODW}=2\Delta / \hbar$ are the frequencies corresponding to Rashba-type SOC and SODW, respectively. In our case, $\rm \bf P$=0, and the initial condition for the spin is $S_k(0)_x$ (the other components of the spin polarization are negligibly small), so we can solve the above
equation analytically in the limit of $\Omega_\text{R}<\Omega_\text{SODW}$ at the Fermi momentum, and get $S_x(t)=S_x(0) \cdot exp (-t/\tau_\text{so})$, where the spin dephasing time is (see for details Supplementary Note 4)
\begin{eqnarray}
\tau_{\text{so}}=\frac{1+\sqrt{1+\Omega_{\text{SODW}}^2 \eta^2}}{\Omega_{\text{R}}^2\eta}+const
\end{eqnarray}
Inserting the exponential dependence of the SODW gap on the excess coverage and keeping all other parameters constant, we get a good fit of the spin dephasing time $\tau_\text{so}$ to the experimental value for a reasonable parameter set of SODW gap $\Delta$=20~meV, $\eta=$1 $\times$10$^{-12}$~s and $ \Omega_\text{R}$=2.6 $\times$10$^{12}$~Hz which is small compared to $\Omega_\text{SODW}$ up to $\delta \Theta$=0.1~ML. Using the value of $\alpha_\text{R}$=1.9~eV \AA~ as the Rashba parameter found in our previous study \cite{Tegenkamp12}, this refers to an extremely small Fermi-wave vector component along the wires ($k_{\text{F},x} \approx$$ \rm 10^{-4}~\AA^{-1}$). This in turn explains the insensitivity of the propagating electrons along the wires against atomic sized defects \cite{Tegenkamp05}.

Focusing on the direction along the wires, the emergent modulated spin texture in real space can be described by dynamic spin spirals which are antiferromagnetically coupled between the wires as illustrated in Fig.~\ref{FIG4}e. The antiferromagnetic coupling along the \per direction is a consequence of the helical nesting found in ARPES \cite{Tegenkamp12}. The spin texture along the direction of wires represents the dynamic analogue to static one-dimensional  skyrmions as observed by STM in magnetic structures \cite{Heinze11,Romming13}. Our results are representative for systems where SOC and electron correlations are of comparable strength and with a negative exchange coupling. While the SODW is robust and insensitive to these couplings the spin polarization is strongly affected and only in the pure SODW phase highly spin-polarized transport can be expected.
In this respect, the continuous improvement in the field of ultra-high spin-resolved spectroscopy will allow us to investigate this new quantum phase and the associated band gap at soonest in more detail \cite{Okuda13, Strocov15}. Our investigations have further shown that the electrons within the SODW phase are strongly correlated with electrons residing as gap states. With a large screened interaction by reducing the excess coverage, the electronic states also become insulating, opening up a possibility for the realization of SODW order induced Luttinger liquid in quantum wires.


\vspace{2ex}

{\bf Methods}\\
{\bf Sample preparation and measurements.} Spin- and angle-resolved photoemission (SARPES) measurements have been performed at the COPHEE end station at the SIS beamline of the Swiss Light Source \cite{Hoesch02}. The data was analysed as described in Ref.~\cite{Meier09}. Atomic wire structures were grown by evaporation of Pb on Si(557) substrates. The Pb-evaporator used in this study has been calibrated before with an accuracy of 0.01~ML by analyzing numerous so-called Devil's staircase phases on Si(111) (1~ML=$\rm 7.83 \times 10^{14}$ atoms per cm$^{2}$) \cite{Hupalo03}.
The morphology of the clean Si(557) substrate as well as the structure of the wire ensemble has been checked by  low energy electron diffraction (LEED) (see Supplementary Note 2). Details about the preparation and coverage calibration are reported elsewhere \cite{Tegenkamp08}.
All photoemission experiments were performed with p-polarized light with a photon energy $h\nu$=24~eV at a base pressure of $\rm 1\times10^{-8}~Pa$.  By recording intensities and spin-induced scattering asymmetries with the two orthogonal Mott detectors for different emission angles, spin-resolved momentum distribution curves (MDCs) close to the Fermi energy have been measured. In order to reveal good signal/noise ratios in reasonable time the spin distribution has been measured 50~meV below the Fermi level. Nearly all ARPES measurements have been performed at $T$=25~K while the excess coverage was deposited during the cooling phase at $\rm 80 \pm 10~K$. Further details about ARPES, STM and the surface (magneto) transport are reported in the Supplementary Notes 1-3.

{\bf Acknowledgements}\\
The financial support by the Deutsche Forschungsgemeinschaft (DFG) through FOR1700  and the Swiss National Science Foundation (Project No. PP00P2 1447421) is gratefully acknowledged.\\

{\bf Author contributions}\\
The experiment was conceived by J.H.D. and C.T. The experiments were performed by C.B., G.L., S.M, J.H.D, and C.T. The SODW formalism was described by T.D. All authors discussed the results. T.D., H.P., J.H.D. and C.T. wrote the manuscript.

{\bf Additional information}\\
Competing financial interests: The authors declare no competing financial interests.


\newpage
Figure captions:

Figure 1:\\[1ex]
{\bf Atomic and electronic structure of quantum wire ensemble.} {\bf (a)} Scanning tunneling microscopy (+1~V) revealing the interwire spacing $d$=1.58~nm of a (223) facet structure (scale bar is 2~nm). The period of 5.8~\AA=$\sqrt3\times\cos(30^{\circ})\times a_{Si}$ correlates with formation of a Pb-$\rm \sqrt3 \times \sqrt 3$ reconstruction (yellow trapeze shows size of the unit cell). The superimposed modulation of $s$=10$\times a_\text{Si}$ is seen as a spot splitting in the LEED patterns in Fig.~\ref{FIG2} as well.
{\bf (b)} Schematic of the band structure in the direction across the wires (\per direction). The two colors denote the spin orientation in each of the subbands. The distance between bands with the \textit{same} spin helicity is $ g\!=\!2 \pi/d\!$=$\rm \!0.4~\AA^{-1}$. The Fermi nesting driven energy gap $ \Delta\!$=20~meV has been determined by ARPES and transport measurements \cite{Tegenkamp05,Tegenkamp08,Czubanowski08}. Right: the equidistant sequence of both spin bands is nicely reflected by the MDC (spin-integrated) taken at the valence band maximum along the \per direction. The spin signature of the subbands has been deduced from spin-resolved measurements \cite{Tegenkamp12}.
{\bf (c)} The giant SOC ensures a constructive interference of Fermi surfaces (FS) with the same helicity in k-space along the \per direction. Renormalization of the FS gives rise to a "hot-spot" wavevector $\Delta k_0$. The direction along the wires is discussed in context with Fig.~\ref{FIG4}a-c.\\[2ex]

Figure 2:\\[1ex]
{\bf Evolution of wire structure with additional coverage.} {\bf (a)} Sequence of the spin-integrated MDCs taken along the \per direction for various excess coverages $\delta \Theta$. The symbols denote the spin texture and have been deduced from spin-resolved measurements (see Supplementary Figure 1). In order to reveal good signal/noise ratios in reasonable time the spin distribution has been measured 50~meV below the Fermi level. {\bf (b)} LEED diffraction profiles along \per direction of the 1.31~ML Pb/Si(557) phase with various $\rm \delta \Theta $. Formation of long-range ordered chains are marked by arrows.  {\bf (c)} 2D LEED pattern with $\delta \Theta$=0~ML  and $\delta \Theta$=0.1~ML. The latter shows nicely the doubling of the period along the \per direction (red arrows) \cite{Czubanowski09}. Optical LEED patterns are shown in the Supplementary Note 2. {\bf (d)} Schematic showing the appearance of new in-gap states. {\bf (e)} The gradual decrease of the band gap $\rm \Delta$ with increasing $\delta \Theta$ deduced from a previous transport measurement \cite{Tegenkamp10}. The error bars deduced from the transport experiments  are around 10~\%.  The (red) regression line follows $\Delta=20~\text{meV} \cdot exp(- \delta \Theta / 0.038~\text{ML})$.\\[2ex]

Figure 3:\\[1ex]
{\bf Damping of spin polarization and coherence length.} {\bf (a)} Spin polarization along the wires measured along the \per direction for different amounts of excess coverage. The spectra are shifted for better visibility ($S_y$-, $S_z$-components and Mott-MDCs are shown in the Supplementary Figure 2). {\bf (b)} Peak-to-peak max-values of all three polarization vectors as a function of excess coverage. The (solid) line is deduced from theory. The dashed lines are guide to the eyes. The error bars are deduced from  least mean square fits to the spin polarized ARPES data. {\bf (c)} Coherence lengths $\xi$=2/FWHM deduced from the FWHM (after deconvoluting the spectrometer function) of the two most intense peaks of the Mott-MDCs. The data point for the clean system was recalculated  from spectra taken at T = 60~K.\\[2ex]

Figure 4:\\[1ex]
{\bf Spin structure along the wires and spin-orbit density wave.} {\bf (a)} $S_x$- and $S_y$-components along the \para direction ($S_z$ is shown in Supplementary Figure 3). {\bf (b)} MDC taken along the direction of the wires without excess coverage. The peak around $k\approx0$ has a fixed spin polarization. {\bf (c)} Quasi Fermi surface including  replica bands from all $\overline{\Gamma}$-points. The labels refer to those shown in {\bf (a)}, {\bf (b)} and Fig.~\ref{FIG1}c. The 10-fold periodicity correlates with the replicas along the \para direction. {\bf (d)} Spin-orbit scattering times as a function of excess coverage deduced from magnetotransport measurements \cite{Luekermann10}. The dashed line is a fit according SODW theory for $ \tau_{\text{SO},x}$, i.e. along the wires. For reference also the scattering time $\rm \tau_{\text{SO},y}$ for direction across the wires is shown. Inset: log scale. {\bf (e)} Real space sketch of antiferromagnetically coupled spin spirals  ($y$-direction) propagating along the wires ($x$-direction). For the sake of simplicity the 10-fold periodicity is not shown.\\

\clearpage

\end{document}